# MAFin: Motif Detection in Multiple Alignment Files

Michail Patsakis[1,2], Kimonas Provatas[1,2], Fotis A. Baltoumas[3], Nikol Chantzi[1,2], Ioannis Mouratidis[1,2], Georgios A. Pavlopoulos[3], Ilias Georgakopoulos-Soares[1,2,*]

[1] Institute for Personalized Medicine, Department of Biochemistry and Molecular Biology, The Pennsylvania State University College of Medicine, Hershey, PA, USA.
[2] Huck Institute of the Life Sciences, Pennsylvania State University, University Park, PA, USA
[3] Institute for Fundamental Biomedical Research, BSRC "Alexander Fleming", Vari 16672, Greece.
[*] Corresponding authors: izg5139@psu.edu

**Abstract**

**Motivation:** Genome and Proteome Alignments, represented by the Multiple Alignment File (MAF) format, have become a standard approach in the field of comparative genomics and proteomics. However, current approaches lack a direct method for motif detection within MAF files. To address this gap, we present MAFin, a novel tool that enables efficient motif detection and conservation analysis in MAF files, streamlining genomic and proteomic research.

**Results:** We developed MAFin, the first motif detection tool for Multiple Alignment Format files. MAFin enables the multithreaded search of conserved motifs using three approaches: 1) by using user-specified k-mers to search the sequences. 2) with regular expressions, in which case one or more patterns are searched, and 3) with predefined Position Weight Matrices. Once the motif has been found, MAFin detects the motif instances and calculates the conservation across the aligned sequences. MAFin also calculates a conservation percentage, which provides information about the conservation levels of each motif across the aligned sequences, based on the number of matches relative to the length of the motif. A set of statistics enable the interpretation of each motif's conservation level, and the detected motifs are exported in JSON and CSV files for downstream analyses.

**Availability:** MAFin is released as a Python package under the GPL license as a multi-platform application and is available at: https://github.com/Georgakopoulos-Soares-lab/MAFin.

**Contact:** izg5139@psu.edu

**Introduction**
The increase in the number of available organismal genomes and of individual human genomes, in part facilitated by multiple international consortia (Rhie *et al.*, 2021; Darwin Tree of Life Project Consortium, 2022; Karczewski *et al.*, 2020; Exposito-Alonso *et al.*, 2020), requires advanced and scalable algorithms to derive the most useful information from the generated sequences.

Sequence relatedness between genomes is an active research area that potentiates the identification of conserved elements, regions undergoing accelerated evolution, and divergent sequences (Bejerano *et al.*, 2004; Alkan *et al.*, 2011; Pollard *et al.*, 2006). Multiple Sequence Alignment (MSA) enables the alignment of two or more related sequences, to quantify sequence similarity. The alignment output includes mutations that differentiate the sequences, such as substitutions, insertions, and deletions incorporated with alignment gaps. Several different sequence alignment algorithms have been developed (Armstrong *et al.*, 2020; Earl *et al.*, 2014) and are implemented to generate phylogenetic trees (Zhang *et al.*, 2018; Liu *et al.*, 2010), perform comparative annotation (Fiddes *et al.*, 2018), infer the evolution of species (Misof *et al.*, 2014), the evolution of proteins (Caetano-Anollés and Caetano-Anollés, 2003), estimate the age of a gene (Dunn *et al.*, 2013), find high confidence transcription factor binding sites (Daily *et al.*, 2011) and identify clinically relevant mutations (Bao *et al.*, 2008), among various applications.

Motifs refer to short patterns in DNA, RNA, or protein sequences that carry biological information (Moeckel *et al.*, 2024). These motifs often reflect functionally important sites involved in regulating gene expression, such as transcription factor binding sites (Georgakopoulos-Soares *et al.*, 2023), and sites associated with protein function and localization (Vazquez *et al.*, 1993). Several different motif detection tools have been developed such as The Meme Suite (Bailey *et al.*, 2015; Grant *et al.*, 2011) and HOMER (Heinz *et al.*, 2010), aiming to find motifs in individual sequences. Additionally, motif databases such as JASPAR (Rauluseviciute *et al.*, 2023), enable the aggregation of biological motifs, often stored in k-mer or as Position Weight Matrices (PWMs) formats, while in certain cases motifs are also stored as regular expressions (Todd *et al.*, 2005). However, even though MSA files are becoming increasingly more utilized across different research problems and applications, no bioinformatics tool enables the detection of motif instances and their conservation directly from MSA files.

Here, we developed MAFin, the first motif detection tool for multiple alignment files. MAFin identifies motifs across multiple sequences by comparing the aligned sequences in MAF format. MAFin takes as input a file of k-mer motif sequences, PWMs, or regular expression patterns and an MSA file in the Multiple Alignment Format. Its output is in JSON and CSV formats containing motif coordinates, similarity vectors, conservation percentages, and multiple related diagrams.

**Materials and methods**
MAFin is implemented as a native Python command line interface program. It accommodates the detection of motifs in MAF files, identifies and stores all motif instances in the alignment file, and produces multiple summary statistics. The motif discovery can be performed for the reference sequence in the MSA file or across all sequences.

MAFin calculates the conservation of the motifs that are identified across the aligned sequences. The process is based on comparing the gapped sequences within the alignment, while still preserving the alignment structure in the MAF file. The comparison results in a similarity vector that matches the true length of the ungapped motif, excluding positions where both the reference and compared sequences have gaps. The conservation percentage for each motif instance is calculated by counting the percentage of matched base pairs. The total conservation percentage for a motif within a block is then estimated by calculating the average conservation percentages of all motif instances in the genomes present in the block.

To guarantee the reliability of MAFin as new features are incorporated, unit tests were automatically executed via GitHub workflows with each pull request. These tests were created using Python's standard library unit testing framework, ensuring that no additional dependencies are introduced while testing the most fundamental features of MAFin.

**Algorithmic process**
MAFin processes the MAF file in chunks, which are distributed across multiple processes to leverage parallel computation. Each process handles a portion of the file, parsing MAF blocks and searching for motifs within the specified genomes (**Figure 1a**).

*k-mer Searching with Aho-Corasick Algorithm:* MAFin leverages the Aho-Corasick algorithm for k-mer searches, a string-searching technique that efficiently processes multiple patterns concurrently in linear time relative to the length of the text. This methodology substantially enhances the speed of pattern matching in large sequences by constructing a finite automaton that encapsulates all k-mers, facilitating the rapid detection of motif occurrences without requiring individual searches for each k-mer.

*PWM Searching*: In PWM searches, the tool determines threshold scores aligned with a specified p-value by sampling random sequences that reflect the background nucleotide frequencies. It then scans the sequences with PWMs, reporting positions that surpass the threshold as motif hits.

*Regular Expression Searching*: For regular expression searches, conventional regular expression matching techniques are used to detect motifs within the sequences.

MAFin requires the following inputs: an MSA file (*--maf_file*), and a motif file, which can be in the form of a file of k-mer motif sequences (*--kmer_file*) or a file with PWM motifs in JASPAR format (*--jaspar_file*) or a set of regular expressions, which are comma-separated (*--regexes*).

Optional inputs for the tool include the *--search_in* parameter, which specifies the source IDs within the MAF file to be scanned for motif matches. By default, this is set to *'reference'*, meaning the search is conducted in the reference genome sequence, but it can be set to a specific source ID to search within that particular genome. Users may also provide a file listing genome IDs via the *--genome_ids parameter*; if this is not specified, all source IDs will be considered for motif conservation analysis. The *--reverse_complement* option enables the inclusion of the reverse complement in the search and can be set to *'yes'* or *'no'*. The number of processes used can be adjusted with the *--processes parameter*. For the JASPAR motif discovery feature, the threshold for motif matching is automatically set to a *p*-value of 1e-4. Lastly, the *--background_frequencies* parameter allows users to specify background nucleotide frequencies for A, C, G, and T, accepted as four floating-point numbers summing to 1. If not provided, uniform background frequencies are assumed.

The outputs consist of JSON and CSV files that detail the coordinates of discovered motifs and their similarity vectors and conservation percentages. The JSON file structures its data using key-value pairs, where each key represents an identified motif. These keys contain information such as the source sequence name and motif coordinates, with motifs detected in gapped and ungapped formats. The values provide conservation data for each identified motif, expressed as a binary similarity vector with length equal to the matched motif , where zero indicates a mismatch and one indicates conservation. Specifically, the JSON file records details such as the source name, chromosome, start and end coordinates, strand, type of motif used (whether PWM, regular expression, or *k*-mer), motif length, motif sequence, gapped and ungapped start and end positions from the MAF file block, conservation statistics, and, for PWM searches, the score, p-value, and false discovery rate.

The Multiple Alignment File (MAF) format is among the most widely used MSA formats (Raney *et al.*, 2024) and is the required format for MAFin. MAF files can be generated from other formats, such as the Hierarchical Alignment (HAL) format, using the hal2maf conversion tool (Hickey *et al.*, 2013).

**Examples**

*Example 1: Perfect match*
In this example, the perfect match case is shown, where the reference sequence and the compared one are identical.
**Motif:** ATCG
**Reference Genome:** A - T C G
**Compared Genome:** A - T C G
In this case, the sequences are the same, with gaps aligned in the same positions. MAFin will compare the ungapped bases to maintain the alignment structure.

**Step-by-Step Comparison:**

| Position | 1 | 2 | 3 | 4 | 5 |
|---|---|---|---|---|---|
| Ref Base | $A$ | $-$ | $T$ | $C$ | $G$ |
| Compared Base | $A$ | $-$ | $T$ | $C$ | $G$ |
| Result | Match | Skip | Match | Match | Match |
| Vector | 1 | | 1 | 1 | 1 |

**Similarity Vector:**
Since we skip the gap in position 2, the resulting similarity vector has a length of 4, matching the true length of the motif (ATCG), and is presented with [1,1,1,1]:

**Conservation Percentage:**
The total positions compared (excluding gaps) amount to 4, and all are matches. Thus, the conservation score is 100%.

**Genomic Coordinates for Matches:**
In addition to the similarity vector, MAFin supplies the genomic coordinates of the motif. If the motif begins at position 1,000 in the reference genome and covers 4 ungapped bases (A, T, C, G), the start and end positions would be as follows: start: 1000, end: 1003. MAFin provides these coordinates for the motif in both the reference genome and the aligned genomes.

*Example 2: Mismatched Sequences with Gaps*
In this scenario, mismatches between the reference sequence and the sequence being compared are analyzed to elucidate the resulting similarity vectors.

**Motif:** ATCG
**Reference Genome:** A - T C G
**Compared Genome:** A T - C G

In this case, the sequences exhibit differences, with gaps positioned at varying locations. MAFin will again attempt to compare the ungapped bases, resulting in a similarity vector of length 4 (corresponding to the length of the motif).

**Step-by-Step Comparison:**

| Position | 1 | 2 | 3 | 4 | 5 |
|---|---|---|---|---|---|
| Ref Base | $A$ | $-$ | $T$ | $C$ | $G$ |
| Compared Base | $A$ | $T$ | $-$ | $C$ | $G$ |
| Result | Match | Mismatch | Mismatch | Match | |
| Vector | 1 | 0 | 0 | 1 | |

**Similarity Vector:**
Once more, the gap at position 2 of the reference sequence is disregarded, resulting in a similarity vector that maintains a length of 4: [1,0,0,1]. It is important to note that the last element is absent.

**Conservation Percentage:**
The total number of positions compared (excluding gaps) is 4, with 2 of those positions being matches. Consequently, the conservation score is 50%.

**Genomic Coordinates for Matches and Mismatches:**
Similar to the first example, MAFin supplies genomic coordinates for the motif. If the motif begins at position 1000 in the reference genome and covers 4 ungapped positions, the coordinates are as follows: start: 1000, end: 1003. These coordinates, together with the similarity vector and conservation score, enable users to readily track the conserved motifs across different genomes.

*3. Example of Reverse Complement Searches*
Searching for a motif on the reverse strand typically requires looking for the reverse complement of the sequence.

**Searching reverse strand through Regex Patterns**
**Reference Sequence:** ATCGGCA
**Regular Expression:** C{2}G ( Two times C followed by G )
Given the complex nature of Regex patterns, it is not feasible to reverse and complement the expression itself. Therefore, MAFin reverses and complements the reference sequence and subsequently searches for the pattern within that modified sequence.

Reverse complement match of CCG in sequence: TGCCGAT results in genomic coordinates 3,5:

| 1 | 2 | 3 | 4 | 5 | 6 | 7 |
|---|---|---|---|---|---|---|
| T | G | C | C | G | A | T |

*Searching reverse strand through K-mers*
**Reference Sequence:** ATCGGCA
**K-mer:** CCG
Searching for a *k*-mer on the reverse strand is considerably simpler. In such a case, MAFin simply reverses and complements the *k*-mer sequence and then searches for it on the original strand.

**Reverse complement K-mer:** CGG is found at genomic coordinates 3,5:

| 1 | 2 | 3 | 4 | 5 | 6 | 7 |
|---|---|---|---|---|---|---|
| A | T | C | G | G | C | A |

**Searching reverse strand through PWMs (JASPAR format)**
PWM:

| Position | 1 | 2 | 3 | 4 |
|---|---|---|---|---|
| A | 0.2 | 0.1 | 0.4 | 0.3 |
| C | 0.3 | 0.5 | 0.1 | 0.2 |
| G | 0.4 | 0.3 | 0.4 | 0.5 |
| T | 0.1 | 0.1 | 0.1 | 0.0 |

To search for a motif on the reverse strand, it is necessary to compute the reverse complement of the PWM. This involves reversing the order of the positions and replacing each nucleotide with its complement. Consequently, the reverse PWM is represented as follows:

Reverse PWM:

| Position | 4 | 3 | 2 | 1 |
|---|---|---|---|---|
| A | 0.0 | 0.1 | 0.5 | 0.2 |
| C | 0.2 | 0.1 | 0.5 | 0.3 |
| G | 0.5 | 0.4 | 0.3 | 0.4 |
| T | 0.3 | 0.4 | 0.1 | 0.1 |

The standard process of PWM search is shown in the following workflow diagram (**Figure 1a**).

**A**

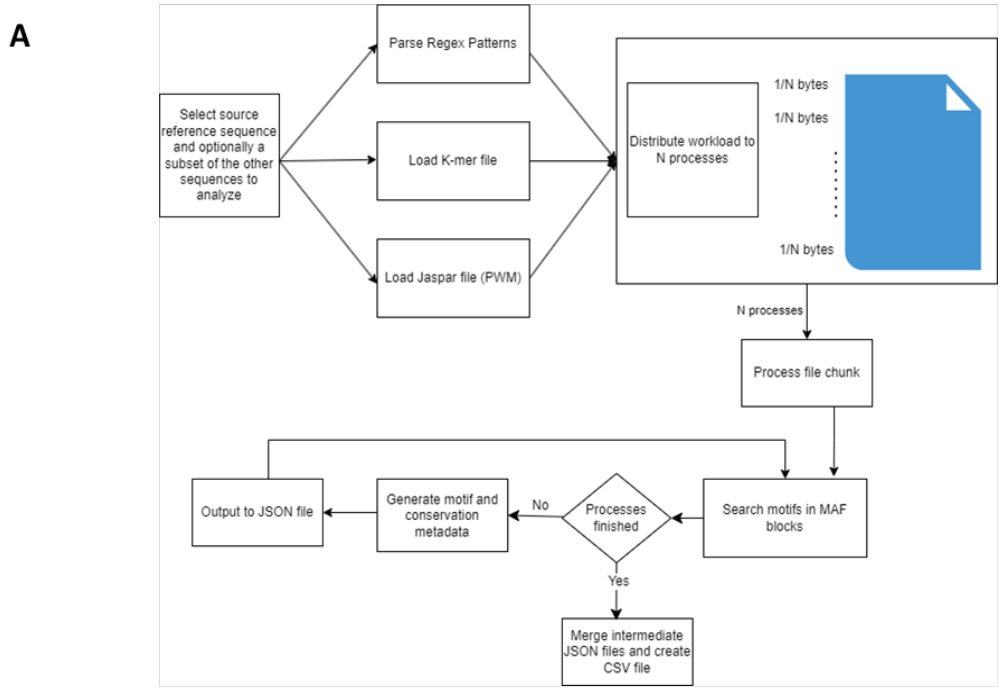

**B**

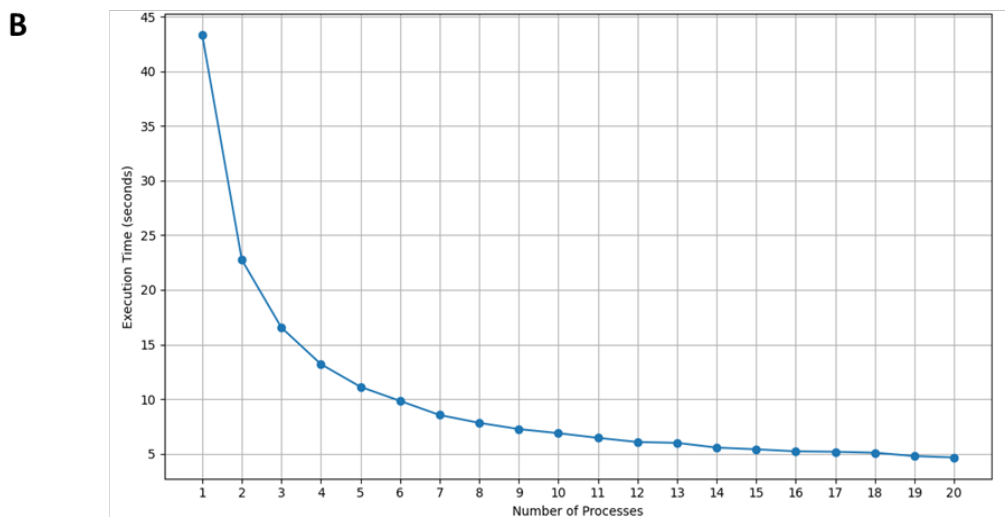

**Figure 1: Workflow and performance of MAFin. A.** Workflow Diagram. This diagram represents the processes and deployment of MAFin. **B.** Execution time relative to the number of processes used.

**Dependencies and Libraries**

The tool is developed in Python and utilizes several external libraries to enhance its functionalities. Specifically, it depends on Biopython for efficient sequence alignment parsing, NumPy for numerical computations, and pyahocorasick for rapid k-mer searching. All of the libraries are open-source and can be easily installed via standard package managers.

**Performance Testing**

A benchmark script was created to perform performance analysis on MAFin when multiple processes are utilized. This script measures execution time as the number of processes increases linearly from 1 to 20 (**Figure 1b**). The results suggest that execution time appears to be inversely proportional to the number of processes.

**Discussion**

In this work, we introduce MAFin, the first motif discovery tool specifically designed for multiple alignments. With the increasing number of sequenced genomes, the use of alignments for comparative analysis is becoming more prevalent, and this trend is likely to continue. MAFin offers a user-friendly and efficient means of discovering motifs within MAF files and assessing the conservation of each motif instance. It generates JSON and CSV outputs for each source sequence name or species in the alignment file, along with summary statistics. Given the growing number of applications utilizing MSA files such as comparative annotation, species identification, phylogenetic tree construction, and functional studies, MAFin facilitates the detection of sequences of interest and functional elements as well as their conservation levels within these alignments.

**Code Availability**

The GitHub code and all the related material is provided at: https://github.com/Georgakopoulos-Soares-lab/MAFin

**Competing interests**

No competing interest is declared.

**Funding**

This work is supported by the National Institute of General Medical Sciences of the National Institutes of Health under award number R35GM155468. F.A.B. and G.A.P. were supported by Fondation Sante; Onassis Foundation; Hellenic Foundation for Research and Innovation (H.F.R.I) under the call 'Greece 2.0 - Basic Research Financing Action, sub-action II, Grant ID: 16718-PRPFOR; Program 'Greece 2.0, National Recovery and Resilience Plan', Grant ID: TAEDR-0539180.

## References

Alkan,C. *et al.* (2011) Genome structural variation discovery and genotyping. *Nat. Rev. Genet.*, **12**, 363–376.

Armstrong,J. *et al.* (2020) Progressive Cactus is a multiple-genome aligner for the thousand-genome era. *Nature*, **587**, 246–251.

Bailey,T.L. *et al.* (2015) The MEME Suite. *Nucleic Acids Res.*, **43**, W39–W49.

Bao,Y. *et al.* (2008) The influenza virus resource at the National Center for Biotechnology Information. *J. Virol.*, **82**, 596–601.

Bejerano,G. *et al.* (2004) Ultraconserved elements in the human genome. *Science*, **304**, 1321–1325.

Caetano-Anollés,G. and Caetano-Anollés,D. (2003) An evolutionarily structured universe of protein architecture. *Genome Res.*, **13**, 1563–1571.

Daily,K. *et al.* (2011) MotifMap: integrative genome-wide maps of regulatory motif sites for model species. *BMC Bioinformatics*, **12**, 495.

Darwin Tree of Life Project Consortium (2022) Sequence locally, think globally: The Darwin Tree of Life Project. *Proc. Natl. Acad. Sci. U. S. A.*, **119**.

Dunn,C.W. *et al.* (2013) Phylogenetic analysis of gene expression. *Integr. Comp. Biol.*, **53**, 847–856.

Earl,D. *et al.* (2014) Alignathon: a competitive assessment of whole-genome alignment methods. *Genome Res.*, **24**, 2077–2089.

Exposito-Alonso,M. *et al.* (2020) The Earth BioGenome project: opportunities and challenges for plant genomics and conservation. *Plant J.*, **102**, 222–229.

Fiddes,I.T. *et al.* (2018) Comparative Annotation Toolkit (CAT)-simultaneous clade and personal genome annotation. *Genome Res.*, **28**, 1029–1038.

Georgakopoulos-Soares,I. *et al.* (2023) Transcription factor binding site orientation and order are major drivers of gene regulatory activity. *Nat. Commun.*, **14**, 2333.

Grant,C.E. *et al.* (2011) FIMO: scanning for occurrences of a given motif. *Bioinformatics*, **27**, 1017–1018.

Heinz,S. *et al.* (2010) Simple combinations of lineage-determining transcription factors prime cis-regulatory elements required for macrophage and B cell identities. *Mol. Cell*, **38**, 576–589.

Hickey,G. *et al.* (2013) HAL: a hierarchical format for storing and analyzing multiple genome alignments. *Bioinformatics*, **29**, 1341–1342.

Karczewski,K.J. *et al.* (2020) The mutational constraint spectrum quantified from variation in 141,456 humans. *Nature*, **581**, 434–443.

Liu,L. *et al.* (2010) A maximum pseudo-likelihood approach for estimating species trees under the coalescent model. *BMC Evol. Biol.*, **10**, 302.

Misof,B. *et al.* (2014) Phylogenomics resolves the timing and pattern of insect evolution. *Science*, **346**, 763–767.

Moeckel,C. *et al.* (2024) A survey of k-mer methods and applications in bioinformatics. *Comput. Struct. Biotechnol. J.*, **23**, 2289–2303.

Pollard,K.S. *et al.* (2006) Forces shaping the fastest evolving regions in the human genome. *PLoS Genet.*, **2**, e168.

Raney,B.J. *et al.* (2024) The UCSC Genome Browser database: 2024 update. *Nucleic Acids Res.*, **52**, D1082–D1088.

Rauluseviciute,I. *et al.* (2023) JASPAR 2024: 20th anniversary of the open-access database of transcription factor binding profiles. *Nucleic Acids Res.*, **52**, D174–D182.

Rhie,A. *et al.* (2021) Towards complete and error-free genome assemblies of all vertebrate species. *Nature*, **592**, 737–746.

Todd,A.K. *et al.* (2005) Highly prevalent putative quadruplex sequence motifs in human DNA.

*Nucleic Acids Res.*, **33**, 2901–2907.

Vazquez,S. *et al.* (1993) Favored and suppressed patterns of hydrophobic and nonhydrophobic amino acids in protein sequences. *Proc. Natl. Acad. Sci. U. S. A.*, **90**, 9100–9104.

Zhang,C. *et al.* (2018) ASTRAL-III: polynomial time species tree reconstruction from partially resolved gene trees. *BMC Bioinformatics*, **19**, 153.